\documentclass[letterpaper]{article} 
\usepackage{aaai24}  
\usepackage{times}  
\usepackage{helvet}  
\usepackage{courier}  
\usepackage[hyphens]{url}  
\usepackage{graphicx} 
\usepackage{natbib}  
\usepackage{caption} 
\usepackage{algorithm}
\usepackage{algorithmic}
\usepackage{pdfpages}

\usepackage{newfloat}
\usepackage{listings}
\DeclareCaptionStyle{ruled}{labelfont=normalfont,labelsep=colon,strut=off} 
\floatstyle{ruled}
\newfloat{listing}{tb}{lst}{}
\floatname{listing}{Listing}

\title{Students' Perceptions and Preferences of Generative Artificial Intelligence Feedback for Programming}
\author {
    Zhengdong Zhang\equalcontrib\textsuperscript{\rm 2}, 
    Zihan Dong\equalcontrib\textsuperscript{\rm 1},
    Yang Shi\textsuperscript{\rm 1}, 
    Noboru Matsuda\textsuperscript{\rm 1}, 
    Thomas Price\textsuperscript{\rm 1}, 
    Dongkuan Xu\textsuperscript{\rm 1} 
}

\affiliations {
    \textsuperscript{\rm 1} North Carolina State University\\
    \textsuperscript{\rm 2} Georgia Institute of Technology\\
    zzhang3135@gatech.edu, \{zdong7, yshi26, nmatsud, twprice, dxu27\}@ncsu.edu
}

\begin{document}

\maketitle

\begin{abstract}

The rapid evolution of artificial intelligence (AI), specifically large language models (LLMs), has opened opportunities for various educational applications. This paper explored the feasibility of utilizing ChatGPT, one of the most popular LLMs, for automating feedback for Java programming assignments in an introductory computer science (CS1) class. Specifically, this study focused on three questions: 1) To what extent do students view LLM-generated feedback as formative? 2) How do students see the comparative affordances of feedback prompts that include their code, vs. those that exclude it? 3) What enhancements do students suggest for improving LLM-generated feedback? To address these questions, we generated automated feedback using the ChatGPT API for four lab assignments in a CS1 class. 
The survey results revealed that students perceived the feedback as aligning well with formative feedback guidelines established by Shute. Additionally, students showed a clear preference for feedback generated by including the students' code as part of the LLM prompt, and our thematic study indicated that the preference was mainly attributed to the specificity, clarity, and corrective nature of the feedback. Moreover, this study found that students generally expected specific and corrective feedback with sufficient code examples, but had diverged opinions on the tone of the feedback. This study demonstrated that ChatGPT could generate Java programming assignment feedback that students perceived as formative. It also offered insights into the specific improvements that would make the ChatGPT-generated feedback useful for students.


\end{abstract}







\section{Introduction}
The rapid recent evolution of artificial intelligence (AI) technology makes the integration of AI into the education sector promising~\cite{chen2020artificial}. Simple AI tools are already widely utilized in education to automate relatively mundane and repetitive tasks~\cite{educause2023}. As AI technology continues to advance in complexity and capability, there is a growing potential to utilize it in more intricate educational tasks. Particularly, Large Language Models (LLMs), such as OpenAI’s ChatGPT, have strong performance on various natural language processing tasks~\cite{chang2023survey}. They present a compelling opportunity for automating a series of natural language-based educational tasks. Specifically, this paper evaluated LLM-generated feedback for programming assignments in an undergraduate Introduction to Java course (CS1) by delving into students' perceptions, using formative feedback guidelines~\cite{shute2008focus} and questions about their preferences of the LLM-generated feedback.

In educational contexts, feedback can be categorized into summative feedback and formative feedback. While summative feedback primarily functions as an evaluative tool at the end of an instructional period, formative feedback offers ongoing guidance aimed at enhancing learning and performance~\cite{dixson2016formative}. Feedback is a crucial component in students’ learning process. Numerous studies across diverse academic disciplines consistently indicate that formative feedback is effective in facilitating student learning~\cite{hao2019investigating}. While the importance of feedback is well-established,  generating feedback for students' programming assignments is challenging since it requires substantial effort by teaching personnel~\cite{gulwani2018automated}. As interest in computer science education continues to surge, the need for automated feedback systems tailored for programming assignments becomes increasingly critical.

Extensive research has been conducted on automated feedback for programming assignments. The state of practice in most classrooms includes evaluating the correctness (e.g.~\cite{morris2003automatic, cheang2003automated}), maintainability~\cite{cordova2021comparison}, and readability~\cite{liu2019static} of students’ code through dynamic analysis and static analysis~\cite{messer2023automated}. AI-based methods, mainly machine learning algorithms trained on various datasets, have also emerged as a complement to traditional methods for programming assignment feedback generation~\cite{piech2015learning,corbett1994knowledge,shi2022code,shi2021toward}.

While these feedback generation systems have been effective in many aspects, there is room for further enrichment in the area of natural language-based automated feedback.  Prior work, such as JavaTutor~\cite{wiggins2015javatutor},  mostly relied on pre-defined templates for natural language-based feedback generation. The rise of LLMs opens up a new opportunity for generating more customized natural language-based feedback without pre-defined templates, as LLMs exhibit strong reasoning and natural language generation abilities~\cite{chang2023survey}. Recent studies also demonstrated the ability of LLMs to solve introductory programming problems~\cite{finnie2022robots} and create new programming problems~\cite{sarsa2022automatic}, suggesting LLMs' potential for generating feedback specific to programming assignments. Given that LLM-generated feedback is an emerging area, there are many research questions that deserve exploration in order to understand how LLMs can be integrated into real-world educational settings, such as how LLM-generated feedback affects learning outcomes and how students perceive LLM-generated feedback.

In this paper, our research focus is on students' perception of LLM-generated feedback. We aim to understand how students perceive the programming assignment feedback generated by ChatGPT, one of the most popular LLMs, in terms of formative qualities. We also seek to understand how students see the comparative affordances of feedback prompts that include their code, vs. those that exclude it since ChatGPT has the capability to offer in-depth suggestions when provided with code. However, there is also a risk of generating inaccurate advice if the code is misinterpreted. Lastly, we gathered students' suggestions for how to further improve the ChatGPT-generated feedback. We conducted our experiments in a CS1 introductory Java course, with 58 students who mostly major in computer science, at a U.S. public university during the Summer of 2023. We designed a series of surveys based on the formative feedback guidelines proposed by Shute~\cite{shute2008focus}. Then we generated automated feedback by ChatGPT API for four selected lab assignments in the Java course. After the grades of each lab assignment had been released, we distributed the feedback to students and collected students' responses through the survey. We also performed a thematic analysis of the open-ended questions in the survey to understand the reasons for feedback preference and the areas for improvement suggested by students. In general, experiments in this paper answered the following three research questions (RQs):

\paragraph{RQ1.}\textit{To what extent do students view ChatGPT-generated feedback as formative?}
\paragraph{RQ2.}\textit{How do students see the comparative affordances of feedback prompts that include their code, vs. those that exclude it?}
\paragraph{RQ3.}\textit{What enhancements do students suggest for improving ChatGPT-generated feedback?}
\vspace{+1mm}

Our study makes several contributions to the field of computer science education. Firstly, by evaluating the ChatGPT-generated feedback against formative feedback guidelines through surveys, we provide empirical evidence that ChatGPT-generated feedback aligns well with formative feedback guidelines established by Shute. Secondly, we find a clear preference among students for feedback generated from prompts that contain their code. Our thematic analysis suggests that this preference is largely driven by the perceived specificity, clarity, and corrective nature of the feedback. Thirdly, this study captures students' suggestions for improving the ChatGPT-generated feedback. According to our thematic analysis, students commonly desire more specific and corrective feedback with sufficient code examples. Additionally, we uncover divergent preferences concerning the tone of the feedback; while some students appreciate an encouraging tone, others opt for more critical evaluations. This divergence underlines the need for personalization in automated feedback systems, offering a direction for future advancements in this area.

\section{Related Work}
\paragraph{Formative Feedback and Feedback Quality Evaluation.}
Formative feedback is generally regarded as effective in educational contexts~\cite{hao2019investigating}. According to Shute, it is defined as “information communicated to the learner that is intended to modify the learner’s thinking or behavior to improve learning”~\cite{shute2008focus}.  Prior research often evaluated the quality of formative feedback by measuring improvement in students’ learning outcomes~\cite{hao2022towards,irons2021enhancing}. This approach, while objective, may not fully capture students' own perceptions of the feedback or the multi-faceted nature of formative feedback as articulated by Shute's formative feedback guidelines. To address this nuanced aspect, we implemented a survey-based methodology, involving students to evaluate the degree to which ChatGPT-generated feedback adheres to Shute's formative feedback guidelines. This methodology enables us to understand students' subjective evaluations of the feedback, thereby enriching the multi-dimensional assessment of formative feedback quality.

\paragraph{Automated Generation of Feedback for Programming Assignment.}

Providing manual feedback for programming assignments is challenging as it requires substantial efforts by teaching personnel~\cite{gulwani2018automated}, and the instructor-to-student ratio is often low for computer science courses~\cite{camp2017generation}. To mitigate this issue, initial efforts focused on automatic grading systems that offered binary feedback on the correctness of code submissions~\cite{von1994kassandra}. The advent of more advanced technologies has enabled the provision of richer and more nuanced feedback. For instance, Singh et al designed a rule-based system to provide feedback on how incorrect a given solution was~\cite{singh2013automated}. Besides, some studies underline the importance of programming styles, developing rule-based tools to assess code style in students' programming assignments~\cite{ala2004supporting}. More recently, AI techniques, such as machine learning and deep learning, have been integrated into the feedback generation process. These technologies are used in intelligent tutoring systems for student modeling~\cite{vanlehn1988student} and undergo modifications to better suit the realm of programming education. For example, Bayesian knowledge tracing~\cite{corbett1994knowledge}, commonly used for tracking student performance in multiple-choice scenarios, was adapted by Shi et al. to be applicable for open-ended programming assignments~\cite{shi2022code}.  Despite these advancements, the potential of ChatGPT in programming assignment feedback generation remains largely unexplored. This paper aims to fill this gap by examining student perspectives on ChatGPT-generated feedback for programming assignments.

\paragraph{Large Language Models for Education.}
The integration of large language models (LLMs), such as ChatGPT, into educational settings, has become an increasing research interest. Many studies were conducted to explore how LLMs can assist students’ learning process, such as generating quizzes~\cite{dijkstra2022reading}, algebra hints~\cite{pardos2023learning}, and code explanations~\cite{macneil2022generating}. Those studies demonstrate that LLMs have the potential to generate educational materials for students.

In a recent study, Maciej et al. explored the potential of using OpenAI's GPT-3.5 model for automated hint generation in programming assignments~\cite{pankiewicz2023large}. They designed a controlled experiment to quantify the impact of AI-generated feedback on students' performance and conducted a quick affect survey to gauge students' emotional states during task completion. The study indicated that compared to the control group, the availability of the AI-generated hints improved students' performance but didn’t significantly impact students’ affective state.

Different from their experimental design that involves students' performance, we mainly focus on understanding students’ subjective evaluation and preference of the ChatGPT-generated feedback and soliciting their suggestions for future improvements. 

Concurrent with the advancements in the educational applications of LLMs, there is a growing concern about the ethical implications of their integration. For instance, Qureshi et al. discuss the possibility of cheating and a shallow understanding of the course material when students are provided with direct access to ChatGPT~\cite{qureshi2023exploring}. These concerns are not isolated. Educational institutions in New York State have banned the use of LLMs due to fears over academic integrity and plagiarism~\cite{chalkbeat}. While the debate over the ethical use of LLMs in education continues, the academic community has started to form guidelines and policies for the use of these tools. Researchers and educational institutions begin to regard LLMs not just as potential sources of academic misconduct but also as useful teaching aids~\cite{apa2023chatgpt,rose2023chatgpt}.

\section{Experimental Design}
\begin{figure}[!t]
    \centering
    \includegraphics[width=1\linewidth]{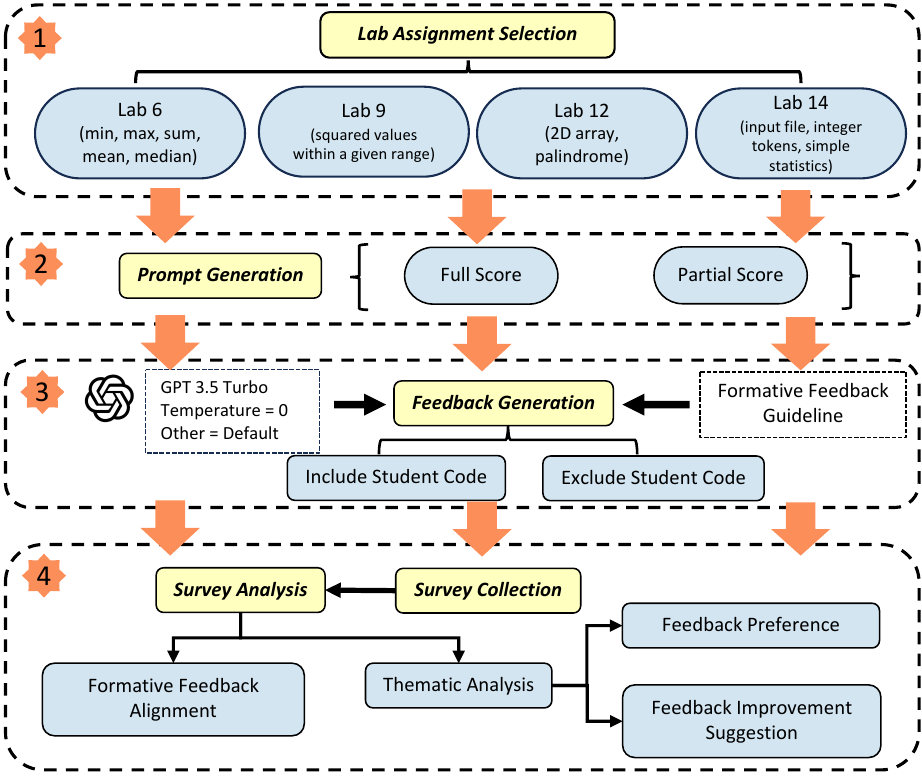}
    \vspace{-2mm}
    \caption{Experimental Design: The Experiment is divided into 4 sequential phases, \textcircled{1}-\textcircled{4}. In \textcircled{1}, we selected 4 lab assignments covering critical topics in the Java course. In \textcircled{2}, we categorized students into groups based on their grades to tailor the prompts accordingly. In \textcircled{3}, Prompts were processed through ChatGPT to generate feedback. And in \textcircled{4}, we collected and analyzed survey results.}
    \vspace{-3mm}
    \label{fig:ExperimentalDesign}
    
\end{figure}

Our IRB-approved experiments were conducted in the Summer of 2023 in an introductory Java programming course (CS1) at a U.S. public university. A total of 58 students consented to participate in the study and completed one or more of our four surveys. The number of respondents for each survey ranged from 23 to 28 students. Our study focused on four specific lab assignments—lab 6, lab 9, lab 12, and lab 14—that were designed to evaluate fundamental Java skills, such as string manipulation, arithmetic calculations, loops, and file processing (see Figure~\ref{fig:ExperimentalDesign} for an overview of our experimental process). These labs were strategically selected to offer a balanced representation of both the course timeline and the varying complexities of the assignments.

All students in the class would receive two pieces of ChatGPT-generated feedback within 1 to 2 days after the professor published the final grades for each of the four selected lab assignments. These pieces of feedback aimed to assist students in improving their grades by evaluating their failed test cases, code, and the GitHub feedback, which was automatically generated by the grading system with each push to GitHub. The first piece of feedback includes students' code in the prompt while the second did not. This design allowed us to explore the impact of including students' code in the prompt on students' preferences towards the ChatGPT-generated feedback. 
After receiving the ChatGPT-generated feedback, students were requested to complete a survey within one week. These surveys were designed to capture students' experiences, satisfaction levels, and suggestions for improvements. This assignment-feedback-survey procedure was repeated for all four lab assignments to study the extent of the formative nature, students' preference for prompts, and potential improvement of ChatGPT in Java assignment feedback generation. 
The procedures for feedback generation and survey generation are elaborated in the sections below.

\subsection{Feedback Generation}
A pivotal part of our experimental design involves the generation of feedback for students' assignments using ChatGPT API. 
This process includes the following key considerations.

\textbf{Message Roles}:
The ChatGPT API allows users to send a list of messages to ChatGPT by two roles: “system” and “user”~\cite{javabydeveloper}. Messages tagged with the “system” role serve as pre-defined instructions that guide ChatGPT's behavior. They are referred to as system prompts in this paper. Messages tagged with the “user” role represent the input from users interacting with the ChatGPT and are termed as user prompts in this paper. Student-specific information such as code and failed test cases was entered into ChatGPT through the “user” role.

\textbf{Temperature Setting}:
Temperature is a parameter in the ChatGPT model that controls the randomness of the model’s outputs. Lower temperature values make the output more deterministic, while higher values introduce more randomness~\cite{openai}. To avoid unnecessary randomness in responses, the temperature was set to 0 for feedback generation in this paper.

\begin{figure}[!t]
    \vspace{-2mm}
    \centering
    \includegraphics[width=1\linewidth]{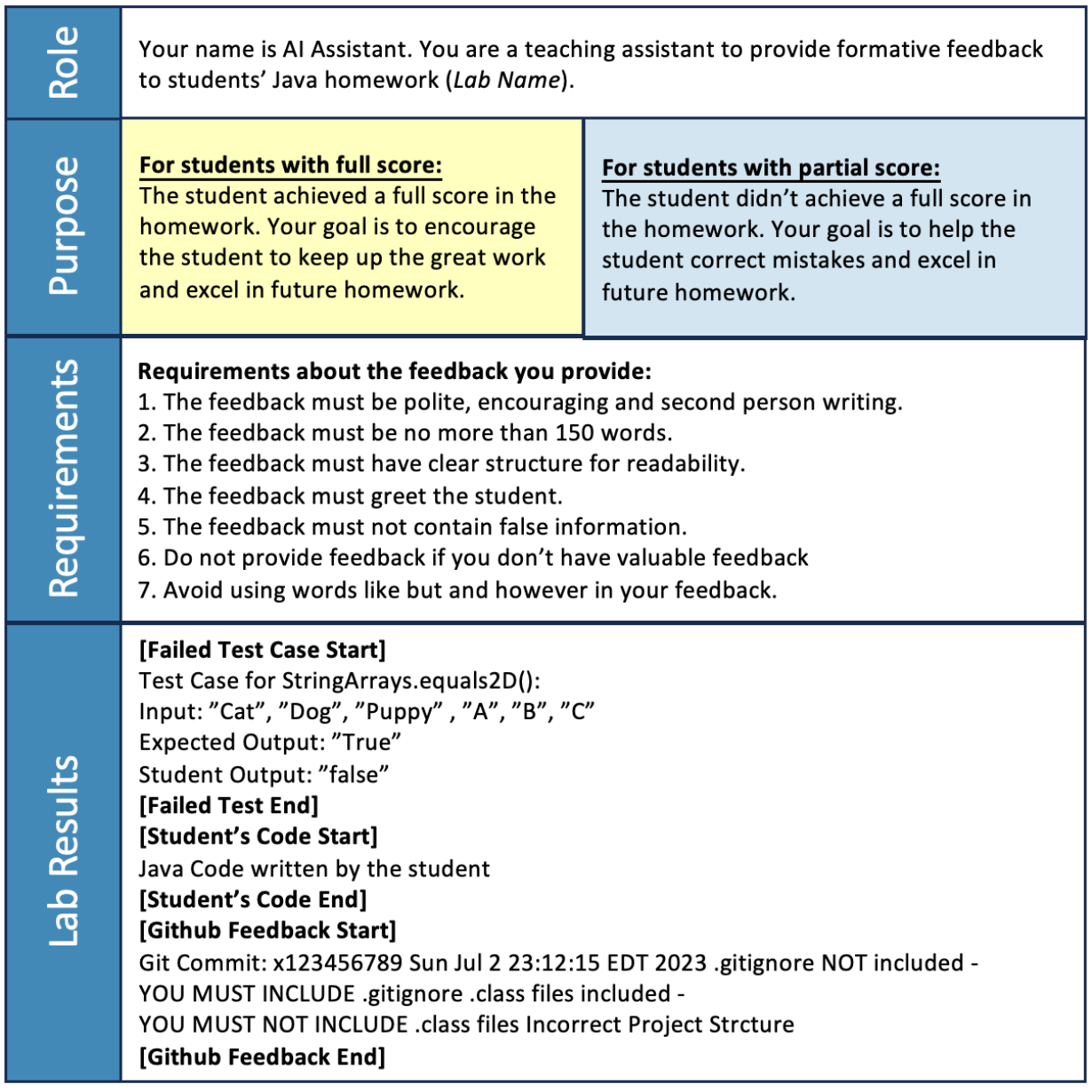}
    \caption{Prompts: In the Role section, ChatGPT was designated as a teaching assistant to provide feedback. The Purpose section categorized students by full or partial scores and tailored prompts for customized feedback. The Requirement section outlined basic criteria for generating feedback, and the Lab Results section included specific student conditions as prompts for feedback.}
    \vspace{-4mm}
    \label{fig: Prompt}
\end{figure}

\textbf{Token Restrictions}:
OpenAI imposes token limits on ChatGPT, which can be challenging when generating feedback for assignments with extensive code. Although our experiments were unaffected by this constraint, we acknowledged it as a potential bottleneck for real-world feedback applications. To assess the impact of including students' code in user prompts on their preferences, we generated two feedback versions for students. One included their code in the prompt, while the other didn't. This approach helped us understand how code inclusion affected perceived feedback usefulness~\cite{openai2}.

\textbf{Prompts}:
In our experiments, prompts were divided into system prompts and user prompts as defined in the Message Roles section. System prompts varied based on students' grading outcomes. For students receiving full scores, the system prompt required ChatGPT to generate words of encouragement and a summary of the student’s performance. For students who did not receive a full score, the system prompt required ChatGPT to give tailored feedback highlighting mistakes in the code and how the student could improve. 
The user prompt was composed of the following components, separated by square brackets indicating the beginning and end of each component (See Figure~\ref{fig: Prompt}): 

\begin{itemize}
  \item Failed Test Case: Details of test cases that students failed, comprising the name of the function these test cases are designed for, input and expected output of these test cases, and the actual output from students.
  \item Students' Code (only included in the version of prompt that contained code): The code submitted by students for an assignment.
  \item GitHub Feedback: the automated feedback generated by the grading system covering build time, Checkstyle errors, commit info, code/test compilation status, presence of .gitignore or *.class files, and project structure validity.
\end{itemize}

\subsection{Survey Design}
To gain insight into students' perceptions and preferences of the ChatGPT-generated feedback, we designed surveys for the selected four lab assignments. Besides the survey agreements, each survey consisted of 10 questions. The first question verified students' performance on the assignment. The next two explored students' preferences regarding the two types of feedback provided. Continuing with six questions, we evaluated students' perceptions of how well the selected feedback aligns with the formative feedback guidelines outlined by Shute~\cite{shute2008focus} for the feedback they received. The last question solicited open-ended comments for future improvement. Shute divided the formative feedback guidelines into four categories: things to do, things to avoid, timing considerations, and learner characteristics. In our study, we concentrated on the first category, as things to avoid, timing considerations, and learner characteristics are not directly relevant to ChatGPT's inherent ability to generate feedback (e.g., one could present ChatGPT-generated feedback immediately, or after a delay). For guidelines within the “things to do” category, we designed survey questions specifically for those requiring students' subjective input. For instance, for the guideline \textit{“Present elaborated feedback in manageable units”}, the survey question is \textit{“Was the feedback presented in small and manageable pieces to avoid overwhelming you?”}, which gauges students’ perspectives of this guideline. The survey questions and associated guidelines can be found in \textbf{Appendix A}.

\begin{figure*}[!t]
    \centering
    \includegraphics[width=0.85\linewidth]{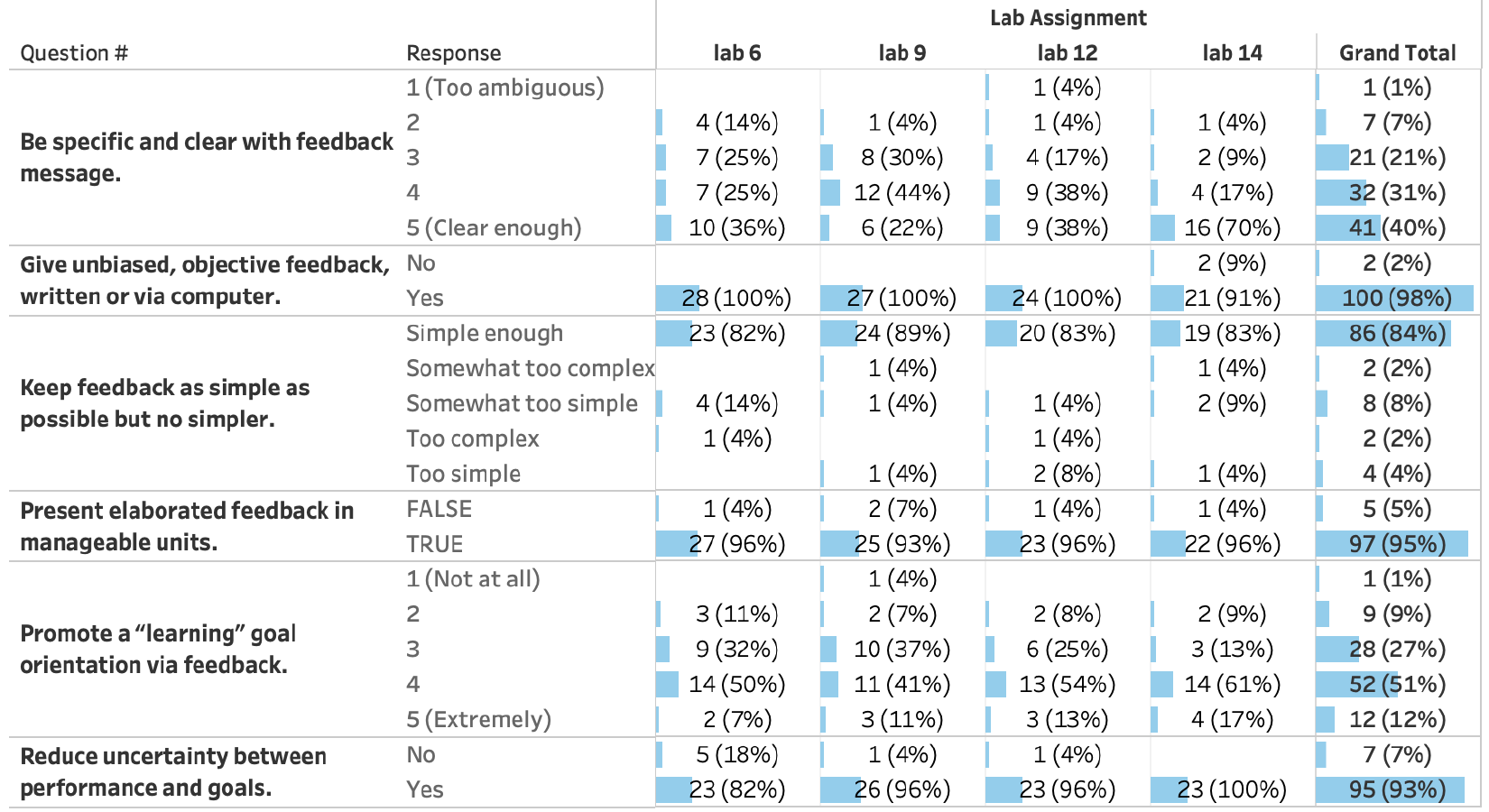}
    \vspace{-2mm}
    \caption{Survey Questions for Formative Feedback Guidelines}
    \label{fig:RQ1}
\end{figure*}

\subsection{Data Processing}
We gathered a range of data to facilitate feedback generation for student assignments. This included students' names which were anonymous by index, code, grades, GitHub feedback, and details of failed test cases. Students' code and GitHub feedback were collected from GitHub/GradeScope using their respective APIs. Information regarding students' grades and repositories was retrieved from the course grading system. Test cases for each assignment were provided by the course lecturer. Utilizing these pieces of data, we ran a specialized Autograder to detect failed test cases for each student. To avoid feeding too many failed test cases to ChatGPT, we categorized failed test cases based on the programming functionality for which the test cases were designed. For students with more than 4 failed test cases, we took only one failed test case from each test case category (if available) and put it into the user prompt. After that, we generated students' outputs for each of these selected failed test cases. The inputs and expected outputs for these failed test cases were extracted from the test case scripts provided by the lecturer. Based on the processed data, we created both system and user prompts for each student. Finally, we leveraged the ChatGPT API to produce tailored feedback.

\subsection{Qualitative Analysis}
We conducted an in-depth analysis of students’ responses to the two open-ended questions in the survey. The analysis involved two key components: categorizing survey submissions based on pre-defined properties and processing open-ended responses through thematic analysis. In the first step, survey submissions were categorized based on conditions including respondents’ grades (either a full or partial score) and their preference for the two pieces of feedback provided. For the thematic analysis \cite{skripchuk2022identifying} in the second step, three authors independently generated initial code to identify key elements in the students' responses, specifically focusing on the reasons behind their feedback preference and suggestions for future improvements. Following this, the researchers collaboratively reviewed and reconciled these initial code to identify overarching themes. These themes were then meticulously refined and explicitly defined (for detailed definitions, see \textbf{Appendix B}).

\section{Results}
In this section, we thoroughly analyzed the survey responses regarding ChatGPT-generated feedback on Java programming assignments. Most notably, we observed a promising alignment between ChatGPT-generated feedback and formative feedback guidelines, with the majority of students offering positive responses. Furthermore, the survey results indicate a significant preference for feedback generated from the prompt that contains the student's code, as it provides more specific, clear, and corrective feedback. Lastly, we captured insights from students on how to improve ChatGPT-generated feedback, offering a direction for future work.

\subsection{Formative Feedback Guidelines Alignment}
We first analyzed the alignment of the ChatGPT-generated feedback with the formative feedback guidelines articulated by Shute. Figure~\ref{fig:RQ1} presents the survey outcomes. The data reveals that the ChatGPT-generated feedback largely aligns with formative feedback guidelines, with over 70\% of students giving favorable evaluations of this alignment.

Among all the guidelines, "Give unbiased, objective feedback, written or via computer." received the highest level of agreement. Overall, 98\% (100/102) of the responses mentioned that the feedback meet this criterion, indicating that most of the students perceive ChatGPT-generated feedback as unbiased and objective.
In addition, 95\% (97/102) of the responses affirmed the guideline "Present elaborated feedback in manageable units." This high rate of agreement can be attributed to our approach of limiting feedback to fewer than 150 words and selectively presenting only a handful of failed test cases, ensuring that the feedback remains succinct and avoiding redundancy for similar test case failures.

Meanwhile, for the guideline "Be specific and clear with feedback message", it is notable that although 71\% (73/102) of the responses viewed the feedback as either somewhat clear or clear enough, only 40\% (41/102) of the responses perceived the feedback as clear enough. This discrepancy underscores the potential for further refinements to enhance the specificity and clarity of ChatGPT-generated feedback.

\begin{figure}[!t]
    \vspace{-4mm}
    \centering
    \includegraphics[width=0.8\linewidth]{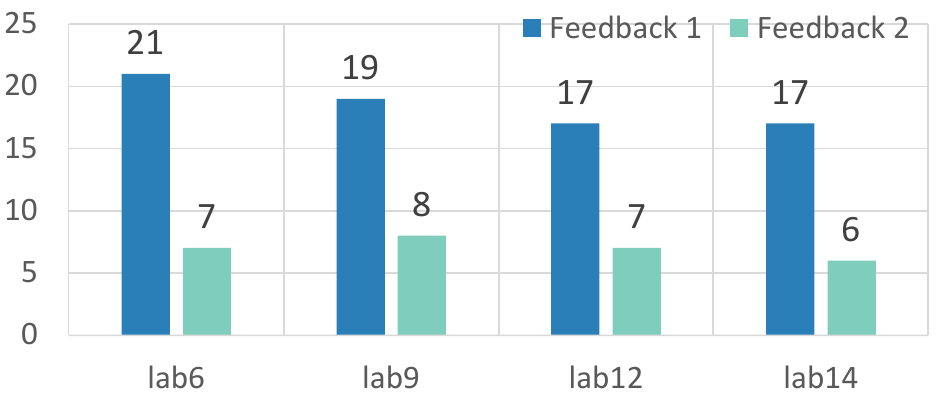}
    \vspace{-4mm}
    \caption{Students' Preference of Feedback }
    \vspace{-4mm}
    \label{fig:Statistics1}
\end{figure}

\begin{figure}[!t]
    \centering
    \includegraphics[width=1\linewidth]{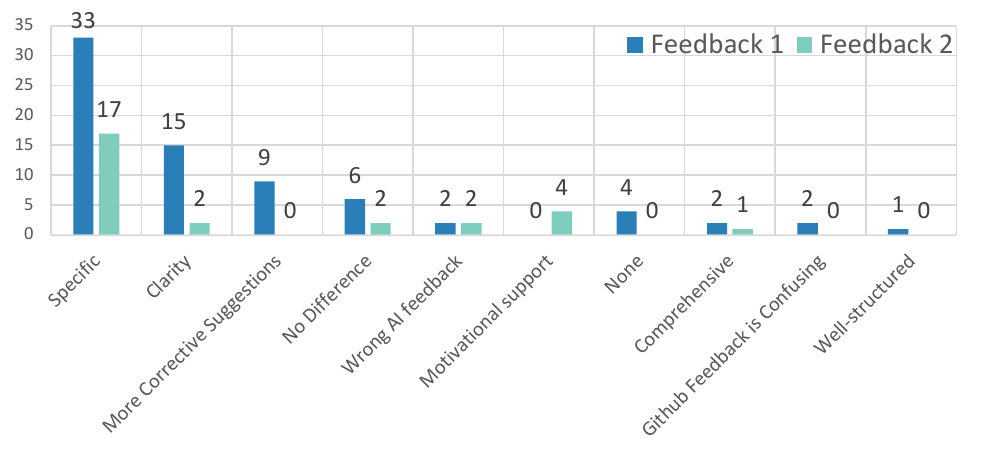}
    \vspace{-8mm}
    \caption{Students' Reasons for Their Preferred Feedback}
    \vspace{-4mm}
    \label{fig:Statistics2}
\end{figure}

\subsection{Impacts of Code Inclusion on Students' Preference}
The survey results indicate that in total 72.5\% (74/102) of the responses preferred the feedback generated with the prompt that contains students’ code (feedback 1) over the feedback without students’ code (feedback 2). This preference is also consistent in each lab (ranging between 70.37\% and 75.00\%), underscoring the significance of including students’ code in the prompt (Figure~\ref{fig:Statistics1}). Intuitively, ChatGPT, when provided with students' code, can offer more specific suggestions on students' code mistakes and thus benefit students more. Through a thematic analysis of the 102 survey responses to the open-ended question asking students why they preferred either feedback 1 or feedback 2, we found that the top three reasons that students preferred feedback 1 were its specificity (33/74), clarity (15/74) and more corrective suggestions (9/74), which accounts for 74\% of the survey responses for feedback 1 (Figure~\ref{fig:Statistics2}). This observation indicates that incorporating code into the feedback enhances ChatGPT's ability to identify specific errors and provide actionable solutions. It also contributes to the readability and clarity of the feedback, possibly because the code snippets make the suggested corrections more concrete than the ones without including the code. 

\begin{figure}[!t]
    \vspace{-3mm}
    \centering
    \includegraphics[width=1\linewidth]{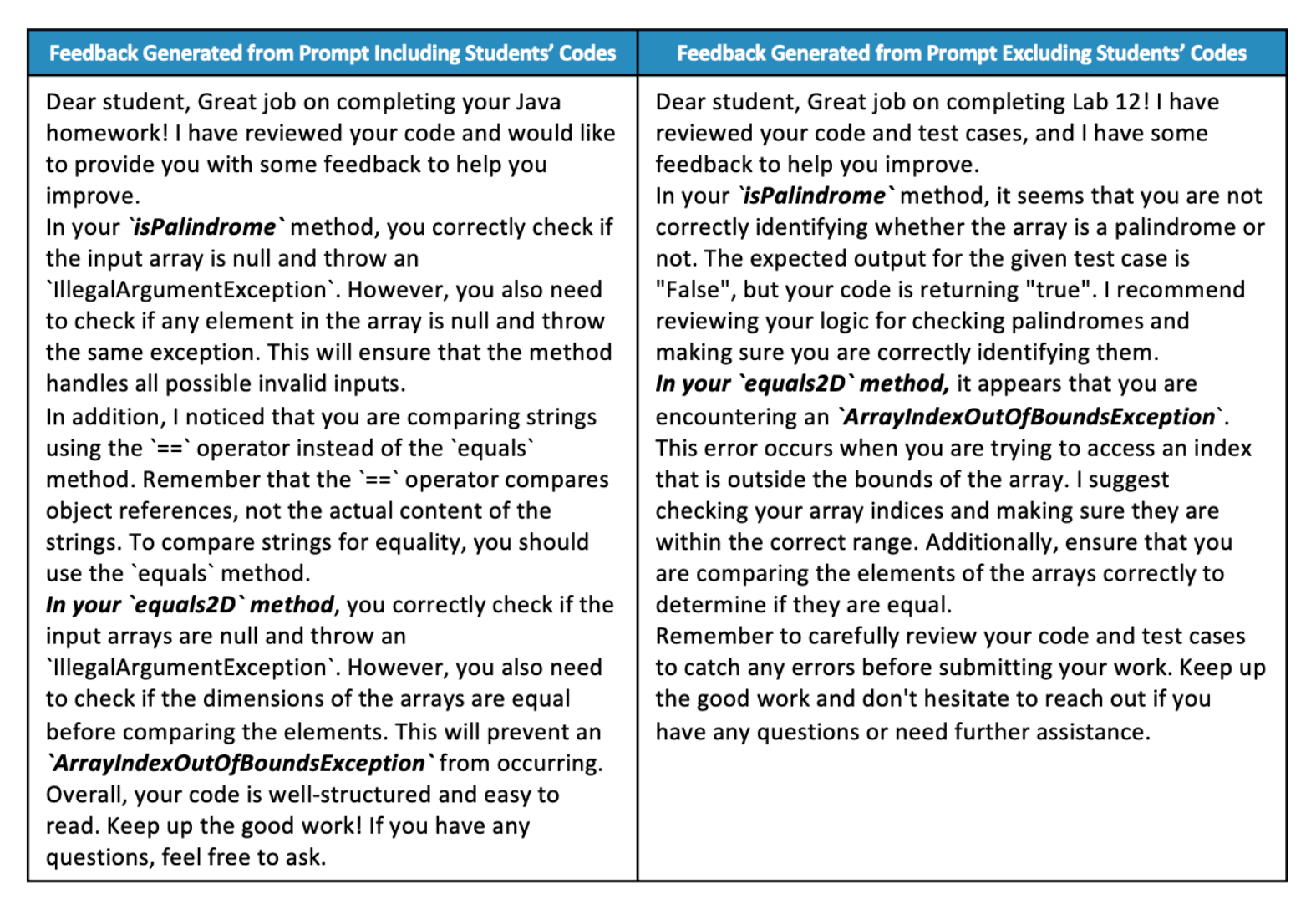}
    \caption{Example of ChatGPT-generated Feedback}
    \vspace{-3mm}
    \label{fig:lab12}
\end{figure}

To illustrate this, we analyzed the entirety of feedback given to students and presented a representative example in Figure~\ref{fig:lab12}, showing the two pieces of feedback provided to a student for the lab 12 assignment. Both pieces of feedback are well-structured, beginning with praising the student for completing the homework and ending by encouraging the student to keep up the good work. 

Nonetheless, feedback 1 excels in its specificity, addressing multiple issues within the student's code, rather than proffering vague critiques. For example, feedback 1 specifically mentions the misuse of the "==" operator for string comparison, whereas feedback 2 mentions a more general issue of the palindrome logic being incorrect without pinpointing the core problem. Furthermore, feedback 1 also provides more explanations. This includes explaining the difference between "==" and "equals" in comparing strings, and why checking array dimensions can prevent the error of "ArrayIndexOutOfBoundsException". In contrast, feedback 2 primarily identifies errors without elucidating their underlying causes or suggesting future improvement opportunities. Consequently, feedback 1, with its more detailed and instructive nature, is favored by students.

While feedback 1 was the clear favorite among the majority, a significant 20\% of participants preferred feedback 2. The primary reasons they cited for this preference were the feedback's specificity (17/28) and its motivational support (4/28). This indicates that while code snippets might make error identification easier, they may also skew the feedback too much toward fault-finding, thus reducing the motivational support provided. Despite the advantages that feedback 2 offers in terms of motivational support, it's still recommended to include code snippets in the prompt for ChatGPT-generated feedback since over 70\% of the survey responses prefer feedback 1 to feedback 2.

\begin{figure}[!t]
    \centering
    \vspace{+1mm}
    \includegraphics[width=1\linewidth]{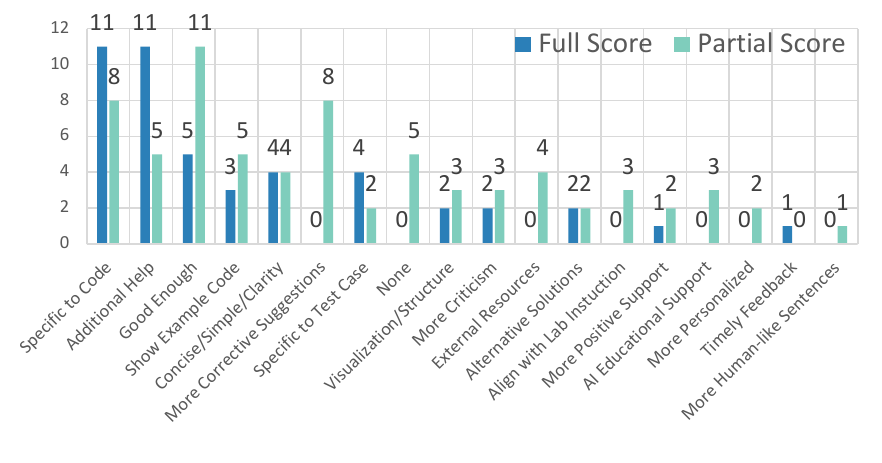}
    \vspace{-8mm}
    \caption{Students' Ideas for Improving ChatGPT-generated Feedback}
    \vspace{-4mm}
    \label{fig:Statistics3}
\end{figure}

\subsection{Improvements Suggested by Students}
In the survey, we included an open-ended question, in which one student could express more than one opinion, to collect students' perspectives on potential improvements to the ChatGPT-generated feedback. A thematic analysis was performed on these responses to identify patterns in students' responses. One major finding from the thematic analysis is that students appreciate the specificity, clarity, and corrective nature of the existing ChatGPT feedback generated from prompts that include students' code, as indicated in Figure~\ref{fig:Statistics2}. However, there is a clear further demand in these areas. Students expressed a particular interest in receiving clearer feedback (8/117), more corrective (8/117), and more specific to their code (19/117) and associated failed test cases (6/117) than the existing ChatGPT-generated feedback. They sought detailed explanations on which test cases their code failed, reasons for the failure, insights into the root causes of issues in their code, and how to fix those issues. 

Furthermore, students advocated for the incorporation of more example code for correct solutions or similar bugs (8/117). As a student suggested, "The AI-generated feedback could include other examples in other programs of bugs similar to mine and why they don't function properly". Students believed this would enhance their understanding of a specific issue or topic and serve as a valuable guide to arrive at the correct solution. In addition, there is a call for receiving external resources (4/117) and additional help beyond the assignment scope (16/117) to supplement students' studies. Interestingly, the desire for such resources varied based on academic performance, as suggested by Figure~\ref{fig:Statistics3}. Students who had achieved full scores were more interested in feedback that extends beyond the course syllabus, while those who had not were more inclined to seek external learning materials and websites.

The analysis also reveals diverging preferences concerning the tone of the feedback. Some students were more receptive to a positive and encouraging tone (3/117), arguing that it enhances motivation. Conversely, others expressed a preference for a more critical tone (5/117) that explicitly outlines mistakes and areas requiring improvement. This finding indicates that a one-size-fits-all approach to feedback tone may not be effective, suggesting the need for personalization to better meet individual preferences.

\section{Discussion}
\paragraph{Teaching Implications.}
In this study, one critical takeaway for educators in computer science is the positive perception of ChatGPT-generated feedback by students, noting its alignment with formative feedback guidelines. This suggests that ChatGPT has considerable potential for generating practical and meaningful assignment feedback. Such automated systems could offer multiple benefits, including improving instructors' efficiency and reducing the financial overhead associated with hiring teaching assistants. Additionally, It's worth mentioning that student preferences for feedback tone varied. Some students favored a positive tone, while others sought more critical assessments. Given these differences, educators could explore personalization options to better meet students' diverse needs in real-world educational contexts. Techniques such as prompt engineering could be utilized to tailor the feedback according to individual preferences.

\paragraph{Research Implications.}
To fully integrate LLMs into the education sector, one practical challenge is to develop interfaces that are readily accessible to educators. A study with European teachers shows that while teachers have a favorable view of AI in education, they seem to have low AI-related skills and a basic level of digital skills~\cite{polak2022teachers}. Thus, one direction of future research is to develop an intuitive software interface that facilitates the end-to-end process from data collection to feedback distribution, enabling educators to efficiently generate, evaluate, and optimize LLM-generated feedback. 
In addition, in our study, students suggest that the LLM-generated could be improved to include code examples and be more specific and clearer. Based on their suggestions, several opportunities exist for enhancing the quality of LLM-generated feedback. Firstly, future research could explore the feasibility of incorporating correct code answers into prompts. This might allow LLMs to identify errors in students’ code efficiently by comparing students’ code with correct code. Secondly, future research can explore the feasibility of fine-tuning LLMs on human-generated feedback and common student errors. By doing so, LLMs might be able to provide more specific and corrective feedback to students. Thirdly, in our study, the LLM model lacks context regarding students' prior actions and their specific needs. Addressing these limitations by improving the model's understanding of students' thought processes might enhance the assistance it can provide.

\paragraph{Limitations.}
The limitations of our study include a small sample size of four labs and 58 students, which might affect the generalizability of our findings. Future work could involve a larger participant pool to validate our results at scale. Additionally, our study mainly focuses on the student perspective of the ChatGPT-generated feedback. Future studies could evaluate LLM-generated feedback from multiple dimensions, including its impact on learning outcomes. Our study is also constrained by the delayed release of grades due to course policy, limiting timely access to feedback. Future research could examine settings where feedback is more immediately available, as this may improve learning outcomes and present different perspectives. Lastly, to evaluate the extent to which students view ChatGPT-generated feedback as formative, we solely look at the absolute percentage of students who provided favorable responses, using a subjective threshold of 70\%. The lack of a baseline for comparison might limit the conclusiveness of our assessment. However, our focus on the student perspective adds an often overlooked but crucial dimension to the evaluation of automated educational feedback systems, offering a meaningful starting point for more comprehensive studies. 

\section*{Acknowledgments}
We would like to express our gratitude to Dr. Sterling McLeod from North Carolina State University for the laboratory support and the reviewers for their thoughtful comments that greatly improved the quality of this paper.

 \newpage
\bibliography{aaai24}

\clearpage


\section*{Supplementary material}

# Appendix A

| Guideline | Survey Questions |
|---|---|
| Present elaborated feedback in manageable units. | Was the feedback presented in small and manageable pieces to avoid overwhelming you? |
| Be specific and clear with feedback messages. | Was the feedback specific and clear enough to be useful for the assignment? |
| Keep feedback as simple as possible but no simpler (based on learner needs and instructional constraints). | Was the feedback content simple enough to understand? |
| Reduce uncertainty between performance and goals. | Was the feedback clear enough to help you understand how well you performed and what you need to do in order to attain the goal of this lab assignment? |
| Give unbiased, objective feedback, written or via computer. | Do you perceive AI-generated feedback as unbiased and objective feedback? |
| Promote a "learning" goal orientation via feedback. | To what extent did the feedback highlight the importance of learning instead of your academic performance in the lab? (e.g. 1. Did the feedback highlight the importance of effort in enhancing learning and performance? 2. Did the feedback underscore the role of mistakes in the learning process? 3. Did the feedback motivate you to prioritize the process of learning?) |

# Appendix B-1

| Label | Definition |
|---|---|
| Specific | AI feedback is specific to the reason why students' code are wrong. |
| Clarity | AI feedback is clear/simple to read. |
| Motivational Support | Ai feedback encourages students to move forward/engage. |
| Wrong AI Feedback | AI feedback can be incorrect sometimes. |
| More Corrective Suggestions | AI feedback explains how to fix a bug. |
| None | No response from students. |
| Comprehensive | AI feedback gives comprehensive feedback that covers wide. |
| No Difference | Whether giving or not giving code to ChatGPT does not make a difference for students. |
| Well-Structured | AI feedback is well-structured, with clear organization in paragraphs. |
| GitHub Feedback is Confusing | GitHub feedback generated from the grading system is confusing. |

# Appendix B-2

| Label | Definition |
|---|---|
| Specific to Test Case | AI feedback explains which test cases are wrong and why they are wrong. |
| Specific to Code | AI feedback is specific to the student's code and show root cause of issues in the student's code. |
| External Resources | AI feedback provides external resources to help students correct issues or further improve themselves. |
| Align with Lab Instruction | AI feedback aligns with assignment instructions. |
| More Criticism | AI feedback gives excessively positive feedback. |
| Show Example Code | AI feedback provides correct assignment code or code of similar problem. |
| Concise/ Simple/ Clarity | The content of AI feedback is not redundant or wordy. |
| Visualization/Structure | AI feedback is well-structured and visually appealing. |
| Good Enough | AI feedback is good enough. |
| More Corrective Suggestions | AI provides more feedback on how to fix issues. |
| Additional Help | AI provides feedback on additional topics that are out of scope of the assignment feedback such as Java style. |
| Alternative Solutions | AI discusses alternative solutions to assignment problem. |
| More Personalized | AI feedback should be more personalized. |
| More Human-like Sentences | AI should generate more human-like sentences in the feedback. |
| None | No response from students. |
| More Positive Support | AI should provide more positive and encouraging feedback. |
| AI Educational Support | Students want to learn more about AI tools and how the AI feedback is generated. |
| Timely Feedback | Students prefer timely feedback. |


\end{document}